\documentstyle[12pt]{article}

\newcommand{\vect}[1]{\mbox{\boldmath $ #1 $}}
\begin{document}

\title{
  Pairing correlation in nuclear matter from Skyrme force
}

\author{
  Satoshi~Takahara, Naoki~Onishi, and Naoki~Tajima\\ 
  Institute of Physics, College of Arts and Sciences,\\ 
  University of Tokyo, Komaba, Meguro, Tokyo 153, Japan
} 

\date{\today}

\maketitle

\begin{abstract}
  The properties of pairing correlation in nuclear matter are
  investigated by using various versions of Skyrme forces.  Truncation
  of states involving pairing correlation, necessary due to zero range
  nature of the Skyrme force, is discussed in detail.  A plateau
  appears in pairing gap versus cutoff for each force. We propose to
  choose the cutoff parameter in the middle of the plateau so that the
  parameterization is independent of nuclides.
\end{abstract}
%---------------------------------------------------------------------
%\newpage

\baselineskip 0.821cm

The mean field theory with the Skyrme force \cite{Sk56,VB72} has been
successful in describing properties of the atomic nucleus. Bulk
features such as binding energies and radii have been reproduced well
and, in addition, more detailed properties such as fission barriers,
giant resonances, single-particle spectra and low-lying coherent
particle-hole and particle-particle excitations have been studied.
Various versions of the force have been constructed for specific
purposes.

Dobaczewski et al.~\cite{DFT84} proposed the SkP force in order to
describe the pairing properties.  While pairing correlation
contributes to various properties of nuclei, one of the most
interesting topics is the formation of skins and halos in neutron
excess nuclei. In these nuclei the neutron Fermi energy rises
close to but below zero so that unoccupied neutron orbitals just above
the Fermi level are unbound.  Therefore the pairing interaction
scatters particle pairs from bound states to continuum states. This
process can have sizable amplitude in the peripheral region and the
formation of neutron skin or halo can be enhanced. Pairing correlation
in neutron excess nuclei is also discussed in other articles, e.g.
ref.~\cite{BE91}.

In the Hartree-Fock (HF)+BCS method, the mean field is solved
self-consistently but the pairing field is separately treated. The
pairing interaction is usually taken to be so called seniority force
or the monopole force. The force is independent of the orbital so that
gap equations are decoupled from the HF field and the whole scheme
becomes extremely simple. For each nucleus, the strength of matrix
element $G$ is determined by, e.g. fitting to experimental odd-even
mass staggering. The matrix element $G$ must, however, be dependent on
quantities other than $N$ and $Z$, e.g., the density and deformation. In
order to deal with the pairing gap more properly, it is desirable to
utilize interactions independent of nuclides like the Skyrme force as
a more fundamental approach.

The treatment of HF+BCS including continuum states may suffer from
unlocalized density due to unphysical ``particle gas'' surrounding the
nucleus.  As shown by Dobaczewski et al.~\cite{DFT84},
Hartree-Fock-Bogoliubov (HFB) theory is free from this difficulty. In
this theory both the mean field and the pairing field are treated on
an equal footing and determined simultaneously from the same
interaction.

Though HFB has been performed with finite range Gogny force
\cite{DG80}, it is practical to employ HFB with the Skyrme force in
describing neutron excess nuclei. In using Gogny force the expansion
in terms of the harmonic oscillator basis provides a poor
approximation to continuum states, even when a large number of
oscillator basis is taken into account.  Thus coordinate-space
treatment is necessary for neutron-excess nuclei.  The Skyrme force
has the possibility of working in coordinate space. This is the reason
why the Skyrme force is employed as a pairing force in describing
nuclei far from stability \cite{DFT84}.
   
The consistent usage of the Skyrme force is also preferable in the
application of Generator Coordinate Method~\cite{TFB92}.

In the present work we investigate the properties of pairing
correlation of Skyrme forces. Firstly, the formalism and notation are
presented.  Secondly, the strength of pairing correlation in nuclear
matter is calculated by making use of various versions of Skyrme
forces. The main subject of this paper is the dependence of the
pairing gap on the cutoff parameter, which is necessary for zero range
forces.

The Skyrme force is usually parameterized as
\begin{equation}
  \begin{array}{rcl}
    V(1,2) &=& t_{0}(1+x_{0}P_{\sigma})\delta
    +\frac{1}{2}t_{1}(1+x_{1}P_{\sigma})(\vect{k}'^{2}\delta
    +\delta\vect{k}^{2})
    +t_{2}(1+x_{2}P_{\sigma})\vect{k}'\cdot\delta\vect{k}\\
    & &\\
    & &
    +\frac{1}{6}t_{3}\rho^{\alpha}(1+x_{3}P_{\sigma})\delta
    +iW_{0}(\vect{\sigma_{1}}+\vect{\sigma_{2}})\cdot\vect{k}'
    \times\delta\vect{k}.
  \end{array}
  \label{sk}
\end{equation}
Following the notation employed in ref. \cite{DFT84}, particle and
pairing density matrices $\rho$ and $\tilde\rho$ are defined as
\begin{equation}
  \rho(\vect{r}\sigma q;\vect{r}'\sigma' q')=\langle\Phi\mid
  a_{\vect{r}'\sigma' q'}^{{\tiny \dag}}a_{\vect{r}\sigma q}
  \mid\Phi\rangle,
  \label{dens}
\end{equation}
\begin{equation}
  \tilde\rho(\vect{r}\sigma q;\vect{r}'\sigma'
  q')=(-2\sigma')\langle\Phi\mid a_{\vect{r}'-\sigma'
    q'}^{{\tiny\dag}}a_{\vect{r}\sigma
    q}^{{\tiny\dag}}\mid\Phi\rangle.
\end{equation}
Owing to the zero range nature of the Skyrme force, the energy
functional depends on only the local properties of the density
matrices.  In HFB theory the energy density ${\cal H}(\vect{r})$ can
be expressed as a function of the particle density $\rho(\vect{r})$,
the kinetic energy density $\tau(\vect{r})$, the pairing density
$\tilde\rho(\vect{r})$ and the pairing kinetic-energy density
$\tilde\tau(\vect{r})$.
These are defined as
\begin{equation}
  \rho(\vect{r})=\sum_{\sigma q}\rho(\vect{r}\sigma q;\vect{r}\sigma q),~~
  \tau(\vect{r})=\vect{\nabla}'\cdot\vect{\nabla}\sum_{\sigma q}
  \rho(\vect{r}\sigma q;\vect{r}'\sigma q)\mid_{\vect{r}'=\vect{r}}~,
  \label{dens2}
\end{equation}
\begin{equation}
  \tilde\rho(\vect{r})=\sum_{\sigma q}\tilde\rho(\vect{r}\sigma q;
  \vect{r}\sigma q)~,~~
  \tilde\tau(\vect{r})=\vect{\nabla}'\cdot\vect{\nabla}\sum_{\sigma q}
  \tilde\rho(\vect{r}\sigma q;\vect{r}'\sigma q)\mid_{\vect{r}'=\vect{r}}~.
  \label{dens3}
\end{equation}
In nuclear matter of isobaric symmetry, Hamiltonian density is given by
\begin{equation}
  \begin{array}{rcl}
    {\cal H} & = & \displaystyle\frac{\hbar^{2}}{2m}\tau
    +(\frac{3}{8}t_{0}+\frac{1}{16}t_{3}\rho^{\alpha}) \rho^{2}
    +\frac{1}{16}\left\{3t_{1}+t_{2}(5+4x_{2})\right\}\rho\tau\\ & &
    \\ & & \displaystyle
    +\left\{\frac{1}{8}t_{0}(1-x_{0})+\frac{1}{48}t_{3}(1-x_{3})
    \rho^{\alpha}\right\} \tilde\rho^{2}
    +\frac{1}{8}t_{1}(1-x_{1})\tilde\rho\tilde\tau~.
    \label{ham}
  \end{array}
\end{equation}
The trial variational function is taken to be of the BCS type,
\begin{equation}
  \mid\Phi\rangle=\displaystyle\prod_{q,\vect{k}} 
  \left\{
  u(k)+v(k)a_{\vect{k}\uparrow q}^{{\tiny\dag}}
  a_{-\vect{k}\downarrow q}^{{\tiny\dag}} 
\right\} \mid 0\rangle~,
\label{BCS}
\end{equation}
where $\vect{k}$ is wave number vector and $k=\mid\vect{k}\mid$.
Then the densities are given by
\begin{equation}
  \rho=\frac{2}{\pi^{2}}\int_{k_{{\rm min}}}^{k_{{\rm max}}}
  k^{2}v^{2}(k)dk +\frac{2}{3\pi^{2}}k_{{\rm min}}^{3}~,~~
  \tilde\rho=-\frac{2}{\pi^{2}}\int_{k_{{\rm min}}}^{k_{{\rm max}}}
  k^{2}u(k)v(k)dk,
  \label{dens4}
\end{equation}
\begin{equation}
  \tau=\frac{2}{\pi^{2}}\int_{k_{{\rm min}}}^{k_{{\rm max}}}
  k^{4}v^{2}(k)dk +\frac{2}{5\pi^{2}}k_{{\rm min}}^{5}~,~~
  \tilde\tau=-\frac{2}{\pi^{2}}\int_{k_{{\rm min}}}^{k_{{\rm max}}}
  k^{4}u(k)v(k)dk.
 \label{dens5}
\end{equation}
In the above equations $k_{{\rm min}}$ and $k_{{\rm max}}$ are
defined in terms of a cutoff energy $E_{{\rm c}}$ as
\begin{equation}
  k_{{\rm min}}^{2}=k_{{\rm F}}^{2}-\frac{2m^{*}(\rho)}{\hbar^{2}}E_{{\rm c}}
  ,~ 
  k_{{\rm max}}^{2}=k_{{\rm F}}^{2}+\frac{2m^{*}(\rho)}{\hbar^{2}}E_{{\rm c}}
  \label{cut}
\end{equation}
where $m^{*}(\rho)$ is the effective mass of nucleon and
$k_{{\rm F}}$ is the Fermi momentum of normal state. In order to
express the pairing gap as a function of density, a constraint
$\rho=\rho_{0}$ is imposed in minimizing the Hamiltonian density
(\ref{ham}).
For this purpose, we utilize the following Routhian,
\begin{equation}
  {\cal H}'={\cal H}-\lambda\rho+\frac{c}{2}(\rho-\rho_{0})^{2}.
  \label{routh}
\end{equation}
In the r.h.s., $\lambda$ is the Lagrange multiplier and the third term
having a positive constant $c$ is included to stabilize the convergence
procedure at negative curvature points. 
If we put $u(k)=\cos\theta(k)$ and $v(k)=\sin\theta(k)$, variation
with respect to $\theta(k)$ yields an equation,
\begin{equation}
  \tan2\theta(k)= -\frac{\displaystyle\frac{\partial {\cal H}}
    {\partial \tilde\rho}
    +\frac{\partial {\cal H}}{\partial\tilde\tau}k^{2}}
  {\displaystyle\frac{\partial {\cal H}}{\partial\rho}
    +\frac{\partial {\cal H}}{\partial\tau}k^{2}-\lambda+c(\rho-\rho_{0})}.
  \label{iter}
\end{equation}
In solving eqs. (\ref{dens4}), (\ref{dens5}) and (\ref{iter}), a na\"
\i ve iteration procedure does not work when the pairing is weak and
special manipulations are necessary to obtain the solution. We
confirmed that the solution actually gives a minimum by giving the
Routhian small variations in $\theta(k)$ and calculating the Routhian.
 
The pairing gap depending on $k$ is defined as
\begin{equation}
  \Delta(k) = - \sum_{\vect{k}'}V_{{\rm pp}}
  (\mid\vect{k}-\vect{k}'\mid) u(k')v(k').
  \label{delta}
\end{equation}
The pair-scattering matrix element $V_{{\rm pp}}(q)$ is of the form,
\begin{equation}
  V_{{\rm pp}}(q)=
  t_{0}(1-x_{0})+\frac{1}{6}t_{3}(1-x_{3})\rho^{\alpha}
  +\frac{1}{2}t_{1}(1-x_{1})q^{2},
  \label{mtrx}
\end{equation}
where a term depending on the angles is omitted because it vanishes
through integration for the angles.

Figure \ref{del_1} shows the pairing gap $\Delta(k=k_{{\rm F}})$ as a
function of the density characterized with the Fermi momentum $k_{{\rm
    F}}$ when $E_{{\rm c}}$=10MeV. This value of $E_{{\rm c}}$ is
chosen so as to correspond to widely used 2$\hbar\omega$ configuration
space.  The results for five parameter sets of Skyrme forces
(SIII\cite{BFG75}, SkM$^{*}$,\cite{BQ82} SGII\cite{GS81},
SkP\cite{DFT84} and SkSC4\cite{ADD92}) are compared.  The term
proportional to $t_{3}$ in eq. (\ref{sk}) is treated as a density
dependent force.  In SIII this term can be dealt with as a three-body
force. In that case, however, pairing correlation is considerably
reduced: the maximum pairing gap decreases by a factor of 1/6 and only
the normal state solution is obtained for density corresponding to
$k_{{\rm F}}\geq$ 0.7fm$^{-1}$.  For comparison the result for Gogny
D1\cite{DG80} force, a finite range force which reproduces well the
pairing properties of nuclei, is taken from ref.  \cite{KRS89}.  The
figure shows: (i) the gap parameter gives the maximum at $k_{{\rm F}}
\sim 0.8$ fm$^{-1}$, which is equivalent to about one fifth of the
saturation density ($k_{{\rm F}}\sim$ 1.33 fm$^{-1}$), (ii) at the
saturation density, with Gogny force the pairing gap of $\sim$ 0.6 MeV
still remains while with Skyrme forces only the normal state solution
is obtained.

One should notice that, as a result of the zero range nature of the
Skyrme force, pairing gap may diverge if the summation in eq.
(\ref{delta}) is carried out toward infinity.  This implies the
necessity of a truncation of particle states for the summation.  Thus
we restricted the range of integration in eqs.  (\ref{dens4}) and
(\ref{dens5}).  Pairing properties of nuclear matter have been
investigated so far in the context of the neutron star.  Jiang and Kuo
\cite{JK88} studied pairing in nuclear matter by making use of Skyrme
forces, but they used the weak coupling approach in which pairing gap
is constant and not zero only in the vicinity of the Fermi surface.
The effect of cutoff on the pairing gap has not been studied yet and
we extensively discuss it in the remainder of this paper.

Figure \ref{e_c} shows the pairing gap for $k_{{\rm F}}=0.8$fm$^{-1}$
as a function of the cutoff parameter $E_{{\rm c}}$. The pairing gap
depends strongly on the cutoff around $E_{{\rm c}}$=10MeV which
is employed in the calculation shown in Fig. \ref{del_1}. At larger
cutoff, however, there is a plateau for each force. The position of
the plateau alters from one parameter to another and seems to be
associated with the momentum transfer $q$ at which the pair-scattering
matrix element $V_{{\rm pp}}(q)$ changes the sign.  Beyond the plateau
the gap starts to increase again.

We should notice that at high momentum transfer the pair-scattering
matrix element becomes positive and hence increasing cutoff leads to
including a large amount of repulsive components in the pairing
interaction. Thus increasing cutoff would not be expected to increase
pairing gap. Nevertheless the calculations show that the pairing gap
keeps increasing together with the cutoff.  It may seem strange at first
sight but can be understood as follows.

Firstly let us explain the mechanism in which increasing cutoff gives
rise to an unlimitedly increasing gap. The $uv$-factors characterizing
the BCS solution are parameterized with $\theta(k)$. For large cutoff
$k_{{\rm max}}$, $\theta(k)$ takes on negative values for $k$ larger
than a certain value (see eq.(\ref{iter})). In consequence, pairing
kinetic-energy density $\tilde\tau$ changes the sign whereas pairing
density $\tilde\rho$ remains negative, because contribution from high
momentum in $\tilde\tau$ is larger than that in $\tilde\rho$. Thus the
term proportional to $\tilde\rho\tilde\tau$ in the energy functional
(\ref{ham}) comes to have the effect of lowering energy,
notwithstanding the fact that this term usually plays the role of
raising energy.

Secondly we consider the physical meaning in such solutions. A similar
circumstance, in which a repulsive interaction contributes to the pairing
correlation, is reported \cite{KRS91} in the Relativistic
Hartree-Fock-Bogoliubov (RHFB) theory in which the force parameters
are determined to simulate the Gogny force. They stated that this can
happen only in cases where the gap parameter as well as the
interaction changes the sign as a function of the momentum.  In the
case of the Skyrme force the gap parameter actually changes the sign
at high momentum and contributes coherently to pairing correlation.
But $V_{{\rm pp}}(q)$ increases rapidly as a function of the momentum
whereas in the case of finite range force $V_{{\rm pp}}(q)$
diminishes.  Therefore the contribution of repulsive part of the force
is fictitious in the case of the Skyrme force although it is probably
realistic for RHFB. The former force is constructed so as to describe
the ground state properties of nuclei, and corresponds to the lowest
order expansion in momentum. Its behavior at large momentum transfer
does not make sense.

If one takes the cutoff in the middle of the plateau, the dependence
of the gap on the cutoff can be made reduced considerably. Thus the
parameterization (including cutoff) is expected to be independent of
nuclides.  However, since the plateau extends over up to 150 MeV the
question arises as to how to determine the cutoff parameter. Here we
propose to determine cutoff such that $\Delta(k_{{\rm max}})=0$.

We also study the dependence of thus-defined cutoff momentum on the
density.  SIII and SkSC4 preserve almost constant value over the whole
density whereas the others, especially SkP, possess notable
dependence.  In the application of HFB to finite nuclei, the pairing
gap contributes most at the nuclear surface.  Therefore we decided to
take the cutoff momentum to be that at $k_{{\rm F}}\sim 0.8$ fm$^{-1}$
where the pairing gap has maximum in Fig.  \ref{del_1}. Table 1 shows
the cutoff momentum and corresponding cutoff energy related through
eq. (\ref{cut}).

Figure \ref{del_2} shows the pairing gap $\Delta(k=k_{{\rm F}})$
calculated with cutoff given in table 1 as a function of the density
for $k_{{\rm F}}$=0.8fm$^{-1}$.  Some features are different from Fig.
\ref{del_1}. The peak is about three times as high as that in fig.
\ref{del_1} for SkP, SGII and SkM$^{*}$, whereas that for SIII and
SkSC4 remains small.  The order of the size of gap changes.  The
position of the peak moves to the lower $k_{{\rm F}}$ (0.6 fm $^{-1}$)
for SGII, SkM$^{*}$ and SkP.
 
In conclusion, we have investigated the pairing properties of the
Skyrme force in nuclear matter on the basis of HFB. In particular, the
dependence of pairing gap on the cutoff is studied. We have observed
the appearance of a plateau in the pairing gap at a certain cutoff
characteristic of each force.  The formation of the plateau is
attributed to the fact that the pair-scattering matrix element changes
the sign at a certain transferred momentum.  We propose to choose the
cutoff parameter in the middle of the plateau, so that the pairing gap
can be made insensitive to the cutoff parameter and the
parameterization is more likely to be independent of nuclides.

%--------------------- references ------------------------------------%

%--------------------- tables ----------------------------------------%

\newpage

\noindent {\bf TABLE}

\newcounter{tabno}

\begin{list}
  {TABLE \arabic{tabno}. }{\usecounter{tabno}
    \setlength{\labelwidth}{2cm} \setlength{\labelsep}{0.5mm}
    \setlength{\leftmargin}{15mm} \setlength{\rightmargin}{0mm}
    \setlength{\listparindent}{0mm} \setlength{\parsep}{0mm}
    \setlength{\itemsep}{0.5cm} \setlength{\topsep}{0.5cm} }
  \baselineskip=0.821cm

\item 
  \label{tab:cut} 
  Cutoff momentum $k_{{\rm max}}$ defined such that
  $\Delta(k_{{\rm max}})=0$ and cutoff energy $E_{{\rm c}}$
  related with $k_{{\rm max}}$ through eq.~(\ref{cut})
  for $k_{{\rm F}}$=0.8fm$^{-1}$.
\end{list}

\begin{table}[h]
  \begin{tabular}{|c|r|r|}
    \hline & $k_{{\rm max}}$[fm$^{-1}$] & $E_{{\rm c}}$[MeV] \\
    \hline
    SIII & 1.547 & 39.05\\
    SkM$^{*}$ & 1.857 & 61.64\\
    SGII & 2.076 & 80.64\\
    SkP & 2.966 & 169.15\\
    SkSC4 & 1.258 & 19.54\\ 
    \hline
  \end{tabular}
\end{table}

%--------------------- figure captions -----------------------------%

\newpage

\noindent {\bf FIGURE CAPTIONS}

\newcounter{figno}

\begin{list}
  {Fig. \arabic{figno}. }{\usecounter{figno}
    \setlength{\labelwidth}{1.3cm} \setlength{\labelsep}{0.5mm}
    \setlength{\leftmargin}{8.5mm} \setlength{\rightmargin}{0mm}
    \setlength{\listparindent}{0mm} \setlength{\parsep}{0mm}
    \setlength{\itemsep}{0.5cm} \setlength{\topsep}{0.5cm} }
  \baselineskip=0.821cm
\item 
  \label{del_1} 
  Density dependence of the pairing gap for Skyrme forces and Gogny D1
  force. Here the cutoff energy is fixed as $E_{{\rm c}}=10$MeV.

\item 
  \label{e_c} 
  The dependence of the pairing gap on the cutoff energy for various
  versions of Skyrme forces. Diamonds indicate the cutoff energy
  determined such that $\Delta(k_{{\rm max}})=0$.

\item 
  \label{del_2} 
  Same as fig.~\ref{del_1} but with the cutoff momentum given in table
  1.

\end{list}


\begin{thebibliography}{99}

\bibitem{Sk56}T. H. R. Skyrme, Phil. Mag. {\bf 1} (1956) 1043.

\bibitem{VB72}D. Vautherin and D. M. Brink, Phys. Rev. {\bf C5} (1972)
626.

\bibitem{DFT84}J. Dobaczewski, H. Flocard and J. Treiner, Nucl. Phys.
{\bf A422} (1984) 103. %SkP

\bibitem{BE91}G. F. Bertsch and H. Esbensen, Ann. Phys. (N.Y.) {\bf
209} (1991) 327.

\bibitem{DG80}J. Decharg\'{e} and D. Gogny, Phys. Rev. {\bf C21}
(1980) 1568.

\bibitem{TFB92}N. Tajima, H. Flocard, P. Bonche, J. Dobaczewski and
P.-H. Heenen, Nucl. Phys. {\bf A542} (1992) 355.

\bibitem{BFG75}M. Beiner, H. Flocard, Nguyen Van Giai and P. Quentin,
Nucl. Phys. {\bf A238} (1975) 29.

\bibitem{BQ82}J. Bartel, P. Quentin, M. Brack, C. Guet and H. -B. 
Hakansson, Nucl. Phys. {\bf A386} (1982) 79.

\bibitem{GS81}N. V. Giai and H. Sagawa, Phys. Lett. {\bf 106B} (1981)
379.

\bibitem{ADD92}Y. Aboussir, J. M. Dearson, A. K. Dutta and F. 
Tondeur, Nucl. Phys. {\bf A549} (1992) 155.

\bibitem{KRS89}H. Kucharek, P. Ring, P. Schuck, R. Bengtsson and M.
Girod, Phys. Lett. {\bf B216} (1989) 249.

\bibitem{JK88}M. F. Jiang and T. T. S. Kuo, Nucl. Phys. {\bf A481}
(1988) 294.

\bibitem{KRS91}H. Kucharek and P. Ring, Z. Phys. {\bf A339} (1991) 23.

\end{thebibliography}
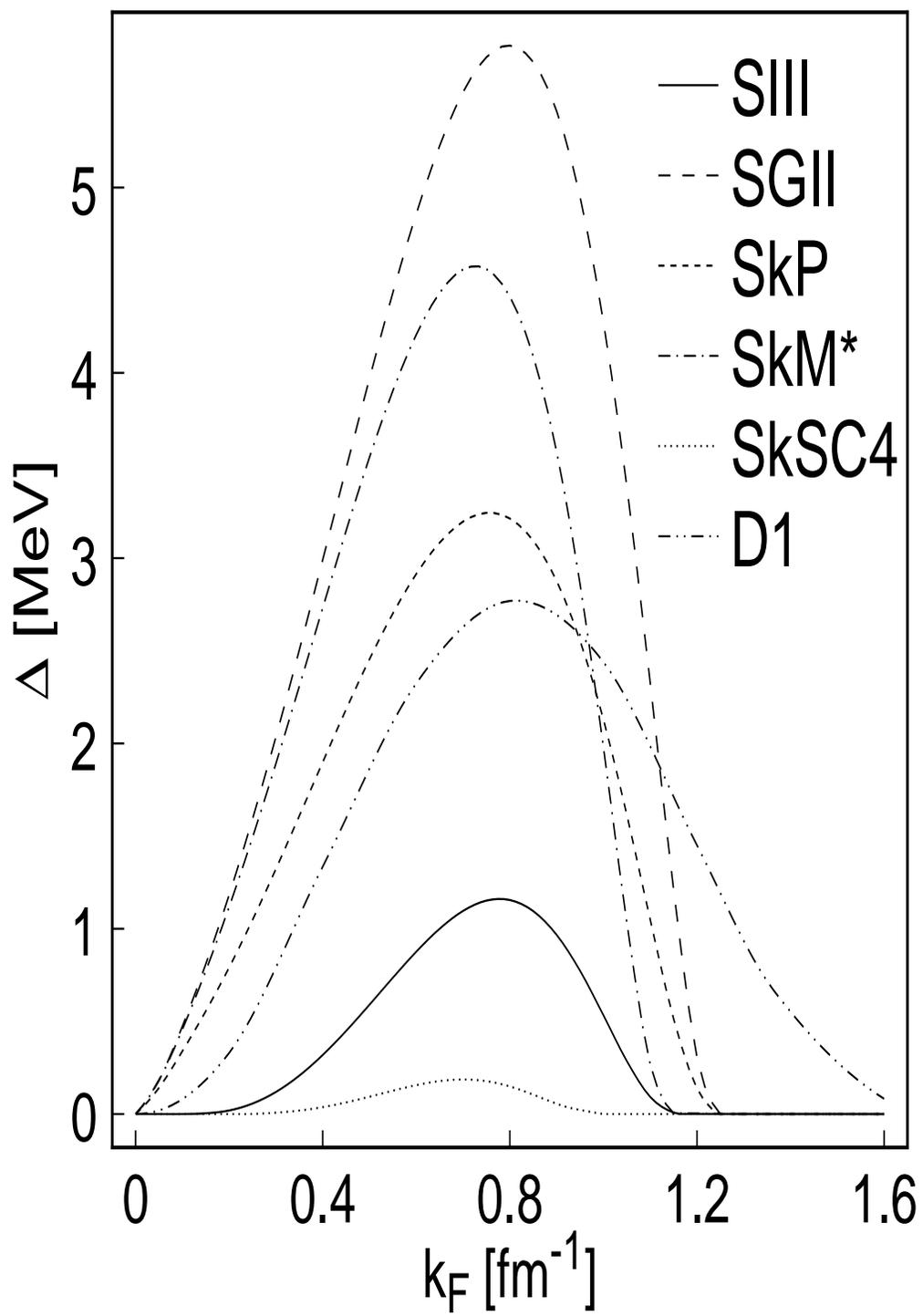
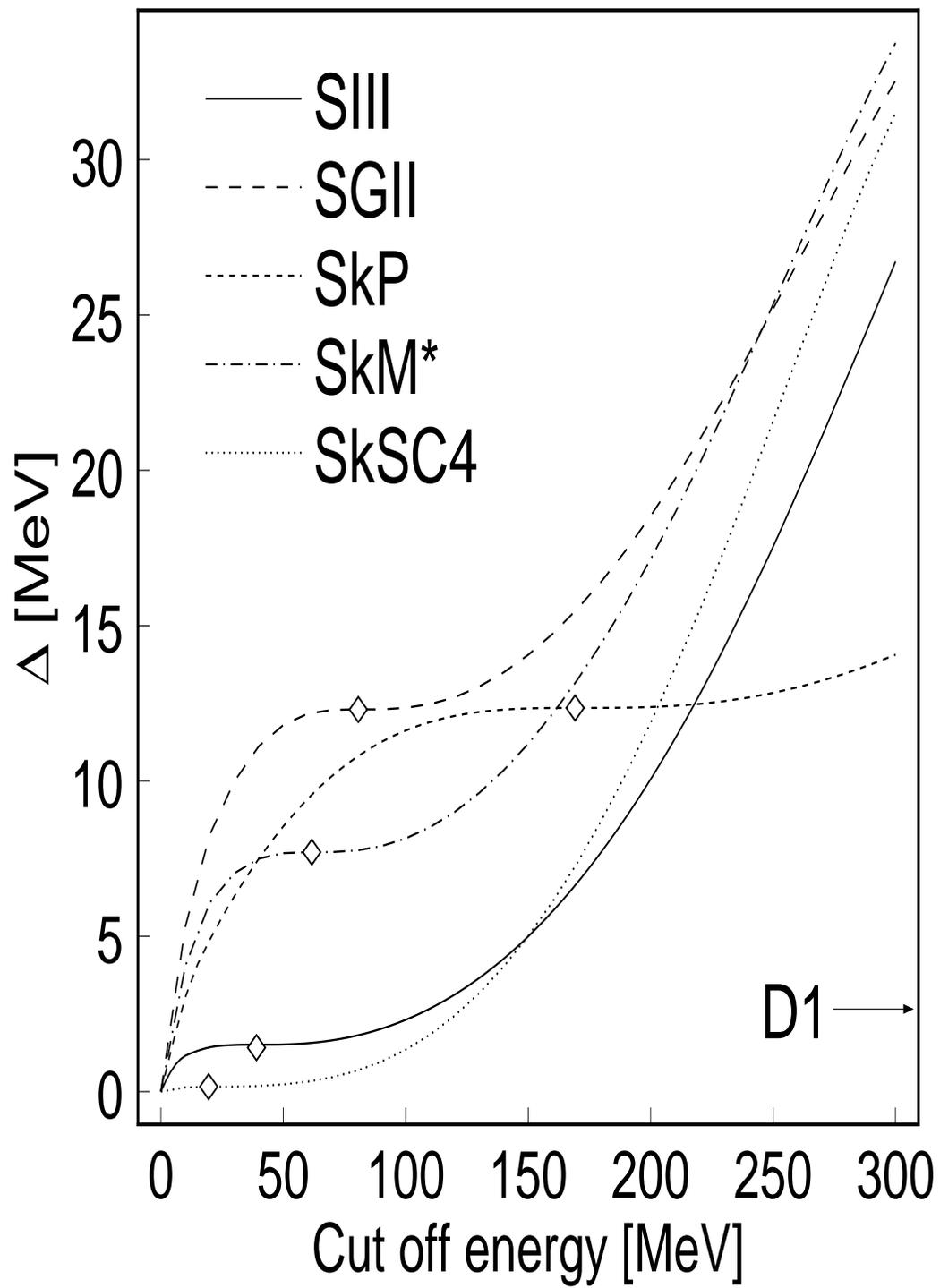
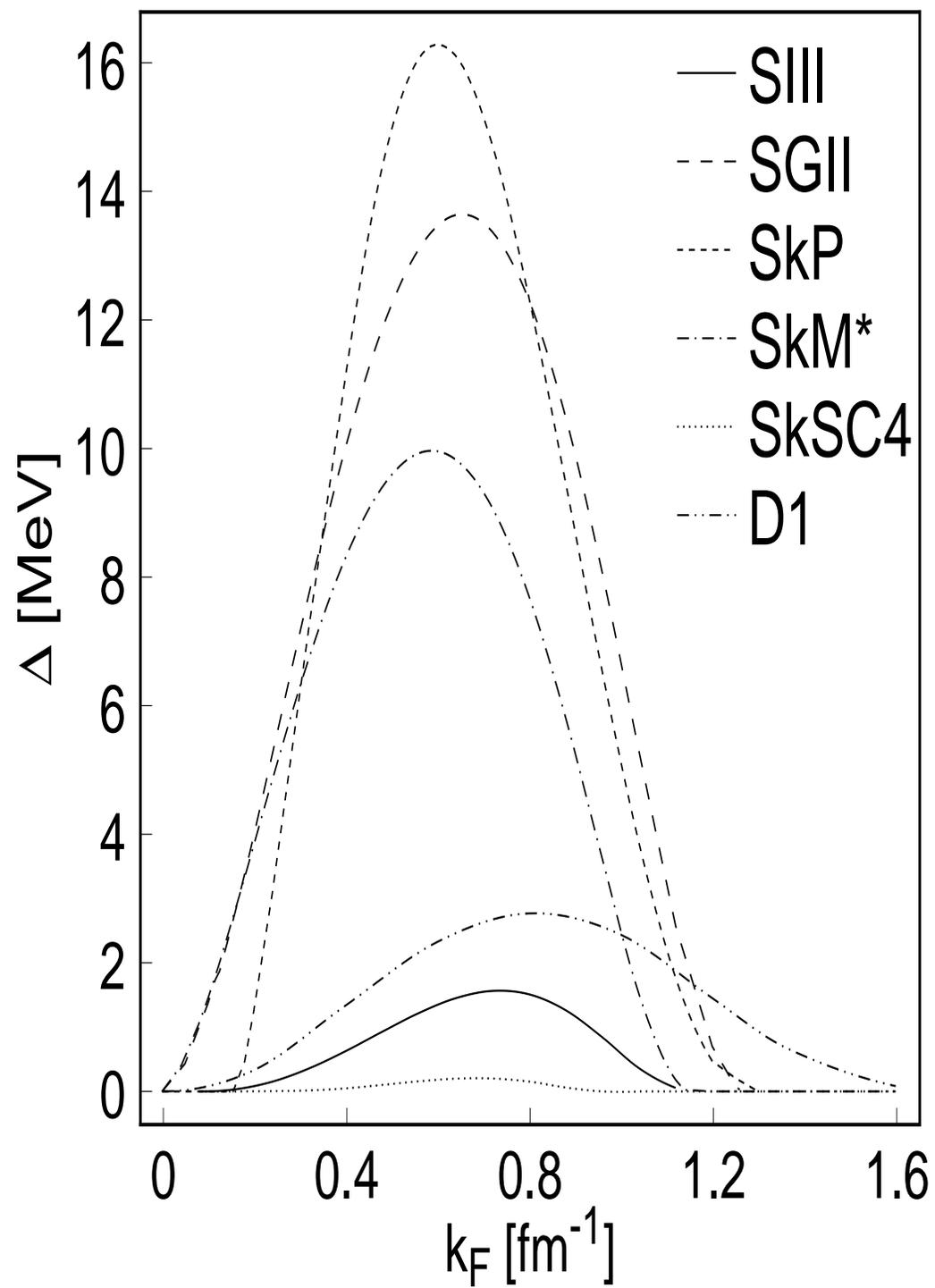
\end{document}